\documentstyle[12pt]{article}
\catcode`\@=11
\@addtoreset{equation}{section}
\def\theequation{\arabic{section}.\arabic{equation}}
\def\section{\@startsection{section}{1}{\z@}{3.5ex plus 1ex minus
   .2ex}{2.3ex plus .2ex}{\bf}}
\def\subsection{\@startsection{subsection}{2}{\z@}{3.25ex plus 1ex minus
   .2ex}{1.5ex plus .2ex}{\bf}}
\def\thesection{\Roman{section}.}

\def\appendixa{ \setcounter{section}{0}\setcounter{equation}{0}
\def\thesection{\bf Appendix A: }
\def\theequation{a.\arabic{equation}}}
\def\appendixb{ \setcounter{section}{0}\setcounter{equation}{0}
\def\thesection{\bf Appendix B: }
\def\theequation{b.\arabic{equation}}}
\def\fph#1#2#3{Fortschr. Phys. #1 (19#3) #2}
\def\jph#1#2#3{Jour. of Phys. #1 (19#3) #2}

\def\nup#1#2#3{Nucl. Phys. #1 (19#3) #2}
\def\phl#1#2#3{Phys. Lett. #1 (19#3) #2}

\def\prd#1#2#3{Phys. Rev. D #1 (19#3) #2}
\def\prl#1#2#3{Phys. Rev. Lett. #1 (19#3) #2}

\def\rmp#1#2#3{Rev. Mod. Phys. #1 (19#3) #2}
\def\tmp#1#2#3{Theor. Math. Phys. #1 (19#3) #2}
\def\zpc#1#2#3{Z. Phys. C #1 (19#3) #2}
\let\al=\alpha
\def\acl{A^{cl}}

\def\aqu{A^{qu}}
\def\aren{A^{ren}}
\def\beq{\begin{equation}}
\def\bq{{\bf q}}
\def\cv{{\cal V}}
\def\dcld{D_{cl}}
\def\dclu{D^{cl}}
\def\dic{\Delta^{cl,-1}}

\def\eeq{\end{equation}}
\def\gcl{G^{cl}}
\def\lcl{\lambda^{cl}}
\def\lqu{\lambda^{qu}}

\def\qr{q^{ren}}
\def\scal{{\cal S}}
\def\scl{{\cal S}^{cl}}
\def\scon{{\cal S}^{con}}
\def\seff{{\cal S}^{eff}}
\def\sgf{{\cal S}^{gf}}
\def\speq{\; = \;}

\def\spmn{\; - \;}
\def\sppl{\; + \;}
\def\squ{{\cal S}^{qu}}
\def\sun{{$SU(N)$} }

\def\tvol{V_{tr}}
\def\vk{\vec{k}}
\def\vkt{\vec{k}_t}
\def\vxt{\vec{x}_t}
\def\zn{{$Z(N)$} }
\begin{document}
\begin{titlepage}
\begin{flushright}
BNL--$\;\;\;\;\;\;\;\;\;$\\ February, 1992\\
\end{flushright}
\vfill
\begin{center}
{\Large \bf \zn interface tension in a hot \sun gauge theory}\\[.1in]
\vfill
{\large \bf Tanmoy Bhattacharya}\\
Service de Physique Th\'{e}orique\\CEN--Saclay\\
91191 Gif--sur--Yvette Cedex, France\\[.2in]
{\large \bf  Andreas Gocksch }\\
Department of Physics\\ Brookhaven National Laboratory\\ Upton, New
York 11973, USA \\[.2in]
{\large \bf Chris Korthals Altes}\\
Centre Physique Th\'eorique au CNRS\\
Section 2, B.P. 907 Luminy \\ F 13288 Marseille, France \\[.2in]
{\large \bf Robert D. Pisarski}\\
Department of Physics\\ Brookhaven National Laboratory\\ Upton, New
York 11973, USA \\[.2in]
\vfill
{\large \bf Abstract}
\end{center}
\begin{quotation}
The interface tension between \zn vacua in a hot \sun gauge theory (without
dynamical fermions) is computed at next to leading order
in weak coupling.
The \zn interface tension is related to the instanton of an effective
action, which includes both classical and quantum terms;
a general technique for treating consistently the saddle points of such
effective actions is developed.
Loop integrals which arise in the calculation
are evaluated by means of zeta function techniques.
As a byproduct,
up to two loop order we find
that the stable vacuum is always equivalent to the trivial
one, and so respects charge conjugation symmetry.
\end{quotation}
\vfill
\end{titlepage}

\section{Introduction}

In the absence of dynamical fermions, \sun gauge theories possess
a global \zn symmetry associated with the center of the gauge group [\ref{r1}].
Confinement implies that at zero temperature the vacuum is \zn symmetric,
but at nonzero temperature there is a phase transition to a deconfining
phase, where the \zn symmetry is spontaneously broken
[\ref{r2}].  In the deconfined phase, a system of infinite extent
falls into one of the $N$ degenerate
vacua, but in a finite volume
bubbles of different vacua form [\ref{r3}], separated by domain walls.
The action of these domain walls is proportional to the interface tension,
and so controls the dynamics of large \zn bubbles.

We previously discussed how
to compute the interface tension $\alpha$ in terms of the
temperature $T$ and the coupling constant $g$ [\ref{r4}].
In weak coupling,
\beq \al \simeq \frac{4 (N-1) \pi^2}{3 \sqrt{3N}} \,
\frac{T^3}{g}
\, \left( 1 - (15.27853...)
\frac{g^2 N}{16 \pi^2} \right)  \, .
\label{e1.1}
\eeq
The term at leading order, $ T^3/g$, is the result of [\ref{r4}];
the principal result of this paper is the computation of $\alpha$ at
next to leading order, $ g T^3$.
In this expression the coupling $g$ represents
a running coupling constant at a temperature $T$; the exact relationship
of this $g$ to the bare coupling constant is given in (\ref{e4.12}).

The physical applications of our results is at best indirect, for it only
applies to a world devoid of dynamical quarks.  It can be
compared to numerical simulations
of lattice gauge theories [\ref{r3}]; indeed,
it was these measurements of $\alpha$
which led us to ask if it is computable in weak coupling.

Nevertheless, the methods and techniques which we have
developed to compute the interface tension
are, we believe, of general interest.
The calculation of the
interface tension reduces to an instanton problem in an effective
theory.  This effective theory is one dimensional [\ref{r5}, \ref{r6}],
as it describes the profile of the interface in the direction
perpendicular to the domain wall.  Remember how
a standard instanton calculation proceeds [\ref{r6}]:
the instanton is the solution
of the classical equations of motion,
with the
action for the domain wall proportional to
$1/g^2$.  The corrections to the classical action are given by expanding
the classical action in a background instanton field.  Corrections
at next to leading order are given by integrating over terms of quadratic
order, which gives corrections to the action of order one.

Contrast this with the interface tension in (\ref{e1.1}).  The action
of the domain wall is $\alpha$ times the transverse volume.  By
its mass dimensionality, $\alpha$ is proportional to $T^3$,
while the constants, and the factors of $N$,
result from the detailed path the \zn interface takes in the space
of \sun gauge fields.  What is surprising is that $\alpha$ starts
out as $1/g$ in weak coupling, for
if the \zn interface were a solution to the
classical equations of motion, it would start instead as $1/g^2$.
This is because
the effective action (in one dimension) which controls
the \zn interface is the sum of the
classical action {\it plus} a quantum term, obtained by
by integrating out fluctuations at one loop order.
In the effective action
the classical piece acts like a kinetic term, and the quantum piece
like a potential.  The \zn instanton is a stationary point
only of the full effective action, and this balance between classical
and quantum terms transforms the usual factor of
$1/g^2$ into a $1/g$.

It is then unclear
how to compute corrections to $\alpha$ beyond leading order.
Surely one cannot compute blindly: if the effective action
is expanded in fluctuations about a background instanton field,
and these fluctuations integrated out, how is
double counting avoided?
That is, how to differentiate
between the quantum fluctuations
which generate the effective action in
the first place, from the quantum fluctuations which are properly
included in the expansion about the instanton?

In this paper we solve this problem for the \zn interface in a manner
applicable to arbitrary
effective actions.  The method generalizes
what is known as the ``constrained''
effective potential [\ref{r7}].  Here we
use it to reduce the four dimensional gauge theory to an effective
scalar theory
in one dimension.  The method is trivial in design:
a delta function constraint for new degrees of freedom is
inserted into the functional integral.  The original degrees of freedom
are then integrated out, producing an effective theory for the new
field.  There is obviously no problem with double counting,
since extra degrees of freedom are introduced in the first place.

For the interface tension, this
method shows that the leading
corrections are of order $g^2$ times that at leading order.
%! Qualification ``of order g^2'' removed
These effects
%!
are due entirely to corrections
in going from the four dimensional theory to
the effective, one dimensional
theory.  They enter both for the kinetic and
potential terms in the effective action.  For example,
the corrections to the kinetic term transform
the bare into a renormalized coupling constant.
%! Last sentence , about fluc's in 1-D, removed.

The order parameter which distinguishes different \zn
vacua is the Wilson line at nonzero temperature [\ref{r2}].
It is convienient to parametrize a vacuum expectation value for
the Wilson line by giving the gauge field $A_0$ a nonzero value;
the effective action for a \zn instanton is then related to the free energy
in such a background field.
The \zn instanton is slowly varying in space,
so that previous results in a constant
$A_0$ field [\ref{r8} -- \ref{r14}],
especially those by Belyaev and Eletsky [\ref{r11}],
Enqvist and Kajantie [\ref{r12}],
and Belyaev [\ref{r14}], can be used.
The technical problem of computing the free energy in
a background $A_0$ field is done most easily by using
zeta function techniques [\ref{r15}].

Much of the interest in considering a hot theory with
$A_0 \neq 0$  concerns the possibility of the vacuum spontaneously
generating an expectation value for $A_0$, and so for the Wilson
line [\ref{r9}--\ref{r14}].  If this happens for $N \geq 3$,
it implies
that the vacuum at nonzero temperature
spontaneously breaks charge conjugation symmetry.
For theories without dynamical fermions,
following Belyaev [\ref{r14}]
we do {\it not}
find evidence for the
spontaneous breaking of charge conjugation symmetry at two loop order:
up to the usual \zn rotations,
the stable vacuum is
the trivial state, with $A_0 =0$ and the Wilson line equal to one.

The outline of the paper is as follows.  In sec. II we review the
calculations of [\ref{r4}].  Sec. III outlines
the general calculation of the interface tension.
The calculation of the interface tension at next to leading order
in Feynman gauge is given in sec. IV.  Sec. V considers the computation
in arbitrary covariant gauges.
There are two appendices.  In appendix A we prove
that the path chosen for the
\zn instanton has minimal action for
$N=3$ and $\infty$.
Appendix B summarizes various integrals
needed in a constant background $A_0$ field.

\section{\zn interface at leading order}

We begin by rederiving the results for the \zn interface tension at
leading order [\ref{r4}].  This is not mere repetition, for here we
use a more natural basis for the generators of \sun matrices than
before.  The subtleties of how to derive
the effective action are deferred until
sec. III.

We work in euclidean spacetime at a temperature $T$, so the euclidean
time $\tau$ varies from $0$ to $\beta = 1/T$.
In the spatial directions the system is a long tube,
of length $L$ in the $z$ direction,
of length $L_t$ in the two remaining spatial directions,
$\vxt$, with $L \gg L_t \gg \beta$.
The volume in
the directions transverse to $z$ is $\tvol = \beta L^2_t$.

A \zn interface is constructed by assuming that the system is in one
\zn phase at one end of the tube, $z=0$,
and in another \zn phase at the other end,
for $z=L$.  This forces a \zn interface along the $z$ direction,
with the action of the interface equal to the interface tension,
$\alpha$, times the transverse volume,
$\tvol$.  As a practical matter, while $L \gg L_t$,
in the end both are taken to infinity.

The \zn symmetry is determined by the
the trace of the Wilson line in the fundamental representation,
\beq
tr \, \Omega(A) \speq \frac{1}{N} \;
 tr \left( {\cal P} \; exp\left( ig \int^\beta_0 A_0(x) d\tau
\right) \right) \; ;
\label{e2.1}
\eeq
${\cal P}$ refers to path ordering.
Let $A_0$ have the value,
\beq
\acl_0(x)\speq \frac{2 \pi T}{g c N} \; q \; t_N \; ,
\label{e2.2}
\eeq
where $t_N$ is the diagonal matrix
\beq
t_N \speq c \;
\left(
\begin{array}{cc}
{\bf 1}_{N-1} & 0      \\
0             & -(N-1) \\
\end{array}
\right)
\;\;\; , \;\;\; c = \frac{1}{\sqrt{2N(N-1)}}
\; .
\label{e2.3}
\eeq
In the presence of this $A_0$ field the Wilson line equals
\beq
tr \, \Omega(\acl) \speq e^{2\pi i q/N} \left( 1 - \frac{1}{N}
\left( 1 - e^{- 2 \pi i q} \right) \right) \; .
\label{e2.4}
\eeq
The trivial vacuum is $A_0 = q = 0$, with $tr \,\Omega = 1$;
\zn transforms of the trivial vacuum occur for
$q = j$, with $tr \,\Omega = e^{2 \pi i j/N}$; $j$ is an integer
between $1$ and $(N-1)$.
Thus the simplest \zn interface is constructed by promoting
the parameter $q$ in (\ref{e2.2}) to a function of $z$, satisfying
$q(0) = 0$ and $q(L) = 1$, so that
$tr \, \Omega = 1$ at $z=0$, and
$tr \, \Omega = e^{2 \pi i/N}$ at $z=L$.

In order to proceed we need a useful parametrization for
the remaining generators of \sun.
The diagonal generators are chosen similarly to (\ref{e2.3}).  For
example,
\beq
t_{(N-1)} \speq \frac{1}{\sqrt{2(N-1)(N-2)}} \;
\left(
\begin{array}{ccc}
{\bf 1}_{N-2} & &\vdots      \\
 & -(N-2) & 0 \\
 \ldots & 0 & 0 \\
\end{array}
\right) \; .
\label{e4.3}
\eeq
Altogether there
are $N-1$ diagonal generators, with $t_i$ from $i=2$ to $N$.
In $SU(2)$ $t_2$ is proportional to the Pauli matrix $\sigma_3$,
while in $SU(3)$ $t_3$ is proportional to the Gell--Mann matrix
$\lambda_8$.

For the off--diagonal generators we follow
Belyaev and Eletsky [\ref{r11}] and use a ladder basis.  For example,
\beq
t^+_{N,1} \speq \frac{1}{\sqrt{2}} \; \left(
\begin{array}{ccc}
\ldots   &  0  & 1  \\
\ldots  & 0  & 0  \\
   & \vdots   & \vdots \\
\end{array}
\right) \; ,
\label{e2.5}
\eeq
\beq
t^-_{N,1} \speq \frac{1}{\sqrt{2}} \; \left(
\begin{array}{ccc}
\vdots & \vdots & \\
0 & 0 & \ldots \\
1 & 0 & \ldots \\
\end{array}
\right) \; .
\label{e2.6}
\eeq
All elements not indicated vanish.
Each diagonal generator $t_i$ has $2(i-1)$ ladder generators
associated with it:
$t^\pm_{i,j}$, with $j$ running from
$j=1$ to $i-1$.  For a ladder generator only one element is nonzero,
\beq
(t^+_{i,j})_{mn} \speq \frac{1}{\sqrt{2}} \; \delta^{i n} \, \delta^{j m}
\;\; , \;\;
(t^-_{i,j})_{mn} \speq \frac{1}{\sqrt{2}} \; \delta^{i m} \, \delta^{j n}
\; .
\label{e2.7}
\eeq
We are working in the fundamental representation, so
the matrix indices $m$ and $n$ vary from $1$ to $N$.

These generators form an orthogonal set, with the diagonal generators
normalized as
\beq
tr(t_i t_{i'}) \speq \frac{1}{2} \, \delta^{i i'} \; ,
\label{e2.8}
\eeq
and the off diagonal generators as
\beq
tr( t^+_{i,j} t^-_{i',j'}) \speq \frac{1}{2} \, \delta^{i i'} \delta^{j j'}
\;\;\; , \;\;\;
tr( t^+_{i,j} t^+_{i',j'}) \speq tr( t^-_{i,j} t^-_{i',j'}) \speq 0 \; .
\label{e2.9}
\eeq
Observe that the metric in the ladder basis is diagonal in the $i$ and $j$
indices, but off-diagonal in the $\pm$ indices.

The advantage of the ladder basis is the simplicity of the commutation
relations.  For the generator $t_N$, since from
(\ref{e2.3}) it involves
the unit matrix in the first $N-1$ components, ${\bf 1}_{N-1}$,
the only nontrivial commutator of
$t_N$ is with its associated ladder
generators, $t^{\pm}_{N,j}$.  This commutator is just
a constant times the same ladder generator:
\beq
[t_N, t^{\pm}_{N,j}] \speq \pm \, cN \, t^{\pm}_{N,j} \; .
\label{e2.10}
\eeq
This relation is familiar from $SU(2)$, from
where up to overall constants, $t_2$ is
$\sigma^3$, and
$t^{\pm}_{2,1}$ are the matrices $\sigma^{\pm}$.

The Wilson line in the adjoint representation can be computed
using (\ref{e2.10}).  This is defined as
\beq
\Omega_{adj}^{a b}(A) \speq
\frac{2}{N^2 -1} \;
\left( t^a \, {\cal P} \; exp\left( i g \int^\beta_0 [A_0(x),]
\right) \, t^b \right) \; ,
\label{e2.11}
\eeq
where ``$a$'' and ``$b$''
refer to the $(N^2 -1)$ adjoint indices, and
$[A_0,]$ denotes the adjoint operator, $[A_0,]X =[A_0,X]$.
For the constant
$A_0$ field of (\ref{e2.2}), the only nontrivial elements of the adjoint
Wilson line are those involving the ladder operators $t^{\pm}_{N,j}$,
and so its trace is
\beq
tr \, \Omega_{adj}(\acl) \speq 1 \spmn
\frac{2}{N+1} \; \left( 1 \spmn cos(2 \pi q) \right) \; .
\label{e2.12}
\eeq
The adjoint Wilson line is unaffected by the \zn symmetry:
for the \zn degenerate vacua, where $q$ is an integer, $tr \, \Omega_{adj} =1$.

The classical action is
\beq
\scl(A) \speq \int^\beta_0 d\tau \; \int d^3 x \;
\frac{1}{2} \; tr \left( G^2_{\mu \nu} \right) \; ,
\label{e2.13}
\eeq
where $G_{\mu \nu} = \partial_\mu A_\nu - \partial_\nu A_\mu
- i g [A_\mu , A_\nu]$ is the field strength tensor.
For the interface problem the field $q$ of (\ref{e2.2}) is assumed
to be a function only of $z$, the length along the tube.   For
this field the classical action reduces to
\beq
\scl(\acl) \speq \tvol \; \frac{4 \pi^2 T^2}{g^2 N}
\; (N-1) \; \int dz   \; \left( \frac{dq}{dz} \right)^2 \; .
\label{e2.14}
\eeq
Using the classical action, a solution to the equations of motion
is $q(z) = z/L$.  There is no true interface,
since the action vanishes like $1/L$
as $L \rightarrow \infty$.  But this is misleading,
for classically there is no sign of the \zn symmetry, as
all values of $q$ degenerate.

This classical degeneracy
is lifted by quantum effects [\ref{r8}].
This is shown by calculating the action in the presence of
the background field in (\ref{e2.2}).
For the time being we assume that $q$ is independent of
$z$, and concentrate on the $q$--dependent terms which lift
the degeneracy in $q$.
With $A_\mu = \acl_\mu + \aqu_\mu$, in background field
gauge [\ref{r16}] the gauge fixing and ghost terms are
\beq
\sgf(A,\eta) \speq \int _0^\beta d\tau \int d^3 x \; \left(
\frac{1}{\xi} \; tr \left(D^{cl}_{\mu}
\aqu_\mu\right)^2 \sppl
\bar{\eta} \left( -D^{cl}_\mu D_\mu \right) \eta \right)\; ,
\label{e2.15}
\eeq
where $D_\mu = \partial_\mu - i g [A_\mu,]$ is the covariant derivative
in the adjoint representation,
$\dclu_\mu = \partial_\mu - i g [\acl_\mu,]$, and $\eta$ is the ghost
field.

For a constant field $q$ it is especially easy to
expand the full action, $\scl + \sgf$, in quadratic
order in the fluctuations $\aqu$, and then integrate them out:
\beq
\squ_1(\acl) \speq \frac{1}{2} \; tr \; ln \left( -\dcld^2 \delta_{\mu \nu}
+ \left( 1 - \frac{1}{\xi} \right) \dclu_\mu \dclu_\nu \right)
- tr \; ln \left(- \dcld^2 \right) \; .
\label{e2.16}
\eeq
The subscript on $\squ_1$ indicates the quantum action at one loop order.
The first term on the right hand side
is from the integration over the gauge fields,
the second from that over the ghosts.  Note that because $\acl_\mu$
is assumed to be independent of spacetime, there are no terms in
the inverse gauge propagator proportional to $G^{cl}_{\mu \nu}$.

This quantum action is independent of the gauge
fixing parameter, $\xi$.  To see this, note that
the derivative of $\squ_1$ with respect of $\xi^{-1}$ is
\beq
\frac{\partial \squ_1(\acl)}{\partial \xi^{-1}} \speq
\frac{1}{2} \; tr \left( - \dclu_\mu \dclu_\nu \left(
\frac{\delta^{\mu\nu}}{-\dcld^2} \sppl \left(1 - \xi\right)
\frac{D^{cl,\mu}D^{cl,\nu}}{(-\dcld^2)^2} \right) \right) \; .
\label{e2.17}
\eeq
That is, the derivative is $- \dcld^\mu \dcld^\nu$ times
the gauge propagator in the background field.  Normally this
propagator is difficult to compute because the covariant derivative
doesn't commute with itself.  For a constant background
field, however, it does,
and so the ordering of the $\dclu_\mu$'s is inconsequential.
Then (\ref{e2.17}) reduces to
\beq
\frac{\partial \squ_1(\acl)}{\partial \xi^{-1}} \speq
\frac{\xi}{2} \;  tr(1) \; .
\label{e2.18}
\eeq
Thus the variation of the quantum action with respect to $\xi$ is
a constant independent of the background field, which can be
dropped.

Adopting Feynman gauge, $\xi = 1$,
the commutation relations
of the ladder basis
reduce the color trace in $\squ_1$ to an abelian
problem.  As $\dclu_0$ is the adjoint covariant derivative, it is
independent of the background field
unless it acts upon the ladder matrices $t^{\pm}_{N,j}$:
\beq
\dclu_0 \, t^\pm_{N,j} \speq \left( \partial_0 \mp 2 \pi T q i
\right) t^\pm_{N,j} \equiv D_{\pm}^0 t^\pm_{N,j} \; .
\label{e2.19}
\eeq
With the euclidean four momentum equal to $(k^0,\vec{k})$, at nonzero
temperature $k^0 = 2 \pi n T$ for integral $n$,
and the covariant derivative becomes
\beq
i D_{\pm}^0 \rightarrow k_{\pm}^0 \equiv 2 \pi T (n \pm q ) \; .
\label{e2.20}
\eeq
The sum over $n$ implicit in the trace includes both positive
and negative values, with the sum over $k^0_-$ equal to that for
$k^0_+$.  Hence the quantum action reduces to
\beq
\squ_1(\acl) \speq 2 (N-1) \;
tr \; ln \left( (k^0_+)^2 + k^2\right) \; ,
\label{e2.21}
\eeq
$k^2 = \vec{k}^2$.

This result is typical of loop effects in a constant
background field.  For the degrees of freedom along the
ladder operators $t^\pm_{N,j}$,
the propagators are as in zero background field, except that
$k^0$ is shifted by a constant amount,
to $k^0_\pm$.  The propagators for the remaining
degrees of freedom are unaffected by the background field.

 From (\ref{e2.21}), at one loop order the
$q$--dependence of the free energy reduces
to $(N-1)$ copies of that for $SU(2)$.
To isolate the $q$--dependence in $\squ_1$, consider its
derivative with respect to $q$:
\beq
\frac{\partial\squ_1(\acl)}{\partial q} \speq
4 (N-1) (2 \pi T)
\; tr \;  \left( \frac{k^0_+}{(k^0_+)^2 + k^2} \right) \; .
\label{e2.22}
\eeq
The integral is most easily done by integrating first over
the spatial momenta:
\beq
(\tvol \, L) \; T \sum_{n = -\infty}^{+\infty}
\; \int \frac{d^3 k}{(2 \pi)^3}
\left( \frac{k^0_+}{ (k^0_+)^2 + k^2} \right)
\speq - (\tvol \, L) \; \pi T^3 \sum_{n = -\infty}^{+\infty}
 (n + q) |n+q| \; .
\label{e2.23}
\eeq
This sum, while formally divergent,
is interpreted using
zeta function regularization [\ref{r15}].
The zeta function $\zeta(p,q)$
is defined as
\beq
\zeta (p,q) \speq \sum_{n=0}^{+\infty} \frac{1}{(n+q)^p} \; .
\label{e2.24}
\eeq
Hence
\beq
tr \left( \frac{k^0_+}{(k^0_+)^2 + \vk^2 } \right) \speq
- (\tvol \, L) \;  \pi T^3 \left( \zeta(-2,q) - \zeta(-2, 1-q) \right) \; .
\label{e2.25}
\eeq
Using
\beq
\zeta(-2,q) \speq - \; \frac{1}{12} \; \frac{d}{dq} \left(
q^2 (1-q)^2 \right) \; ,
\label{e2.26}
\eeq
integration of (\ref{e2.22}) gives
\beq
\squ_1(\acl) \speq \tvol \; \frac{4 \pi^2 T^4}{3} \;
(N-1) \; \int dz \;  q^2 (1-q)^2 \; .
\label{e2.27}
\eeq
{}From the nature of the sum in (\ref{e2.23}),
$\squ_1$ is periodic in $q$, and is invariant under shifts of
$q \rightarrow q + l$ for any integer
$l$.  Thus in (\ref{e2.27}) $q$ is defined modulo one.
Also, a $q$--independent constant in
(\ref{e2.27}) was dropped; this
constant is just the free energy of an ideal gas of $N^2-1$
gluons at a temperature $T$.

As promised, the classical degeneracy in $q$ is lifted by
quantum effects: the minima of the theory are now at
integral values of $q=j$, where
$tr \, \Omega = exp(2 \pi i j/N)$.

In (\ref{e2.27}) the
length in the $z$ direction, $L$, is replaced by the integral
over $z$.  Of course for a constant
field this substitution doesn't matter, but consider the
interface problem.  Introducing the dimensionless coordinate $z'$,
\beq
z' \speq \sqrt{\frac{N}{3}} \; g T  \; z \; ,
\label{e2.28}
\eeq
the sum of the classical and quantum actions becomes
\beq
\scl(\acl) \sppl \squ_1(\acl) \speq
\tvol \;
\frac{4 \pi^2 (N-1)}{ \sqrt{3 N} } \; \frac{T^3}{g}
\; \int dz' \; \left( \left(
\frac{dq}{dz'} \right)^2 + q^2 (1-q)^2 \right) \; .
\label{e2.29}
\eeq

We can view
minimization of this effective action
as a problem in
mechanics, with the coordinate $z'$ as the ``time''.
The classical action contributes the kinetic energy,
$(dq/dz')^2$,
while the quantum action produces a standard double
well potential,
$q^2(1-q)^2$.
For any solution to the equations of motion
the energy,
${\cal E} = (dq/dz')^2 - q^2(1-q)^2$ is conserved, $d{\cal E}/dz' = 0$.
With the \zn interface we want a solution which obeys the boundary
conditions $q(0) = 0$,
$q(L') = 1$, as $L' = (\sqrt{N/3})\,  gT \,
L \rightarrow \infty$ [\ref{r5},\ref{r6}].
By the boundary conditions
the instanton has zero energy, ${\cal E}=0$.
Consequently, for the \zn instanton
\beq
\int dz' \; \left( \left(
\frac{dq}{dz'} \right)^2 + q^2 (1-q)^2 \right) \speq
2 \; \int_0^1 \; dq \; q (1-q) \speq \frac{1}{3} \; .
\label{e2.30}
\eeq
If the total action of
the interface is the transverse volume $\tvol$ times the interface
tension $\alpha$, then at leading order [\ref{r4}]
\beq
\alpha \speq
\frac{4 \pi^2 (N-1)}{3 \sqrt{3N} } \; \frac{T^3}{g} \; .
\label{e2.31}
\eeq

Two assumptions must be justified.  The first is why
$\squ_1$ can be computed for a constant field $q$, and then applied
to the \zn instanton, where $q$ is clearly a function
of $z$.  The reason can be
seen from the definition of $z'$ in
(\ref{e2.28}).  As there is no length scale in the rescaled action,
in terms of $z'$ the instanton's width is of order one.
For the original coordinate, $z$, this implies that the
\zn instanton is ``fat'', with a width of order $1/(gT)$.
In weak coupling this is much larger than the
natural length scale in
a gas of massless, nearly ideal gluons, which is $1/T$.
Thus while the \zn instanton field is large in
magnitude, $\acl \sim
T/g$, it varies slowly in space, and at leading order this variation
can be neglected.  Corrections to this approximation
do enter beyond leading order.

The second assumption is whether in the space of \sun gauge fields
the path chosen is of minimal action.
The \zn instanton interpolates
between $tr \, \Omega = 1$ and $tr \, \Omega = exp(2 \pi  i/N)$.
By a global gauge rotation $\Omega$ can be chosen as
a diagonal matrix, involving the $(N-1)$
diagonal generators of \sun, the $t_i$'s.
The quantum action for a general (constant) field can
be computed directly [\ref{r8}].
We chose the simplest path possible --- straight along
the $t_N$ direction --- but it is not obvious that other
paths, which wander off into the direction of the other
$(N-2)$ $t_i$'s, might not have lower action.  (Our path
is at least a local minimum.)
For $SU(2)$ there is only one path possible.
In appendix A we show
that for $N=3$ and $N=\infty$, the path along $t_N$ is minimal.
On this basis we conjecture
that this remains true for all $4 \leq N < \infty$.

\section{General analysis of the \zn interface}

The partition function of an \sun gauge theory is
\beq
Z \speq \int \; [dA_\mu(x)]
\; [d\eta(x)]\; e^{- \scl(A) - \sgf(A,\eta) } \; ,
\label{e3.1}
\eeq
where $\scl$ and $\sgf$ are the classical and gauge fixing actions of
(\ref{e2.13}) and (\ref{e2.15}).  The gauge field
$A_\mu = \acl_\mu + \aqu_\mu$;
for the time being the choice of $\acl_\mu$ is left open.
The coordinate of four dimensional spacetime is $x=(\tau,\vxt,z)$

In order to reduce this four dimensional theory to an effective
theory in one dimension we introduce the field
\beq
\bq(z) \speq  \frac{1}{\tvol} \; \int^\beta_0 d\tau
\; \int d^2 x_t \;\; \frac{g c N}{2 \pi T}\; \acl_0(x) \; .
\label{e3.2}
\eeq
This choice is obviously motivated by the definition of $\acl_0$;
unlike the scalar field $q$ of (\ref{e2.2}), $\bq$
is a matrix valued field in the adjoint representation
of \sun.  The only subtlety in the introduction of $\bq(z)$ is our
insistence on defining it not just as the average
of $\acl_0$ over time, but over the spatially transverse
directions as well.
The reason for this will become apparent shortly.

The field
$\bq(z)$ is introduced
into the functional integral by a delta function constraint:
\beq
Z \speq \int \; [dA_\mu(x)]
\; [d\eta(x)]\; [d\lambda(z)] [d\bq(z)] \;
e^{- \scl(A) - \sgf(A,\eta) - \scon(\lambda,\bq,A)} \; ,
\label{e3.3}
\eeq
\beq
\scon(\lambda,\bq,A) \speq \int dz \; 2i \; tr \left(
\lambda(z) \left(\bq(z) \spmn
\frac{1}{\tvol} \; \int^\beta_0 d\tau
\; \int d^2x_t \; \frac{g c N}{2 \pi T}
\; \acl_0(x) \right) \right)\; .
\label{e3.4}
\eeq
The constraint field $\lambda(z)$ is introduced to enforce the
definition of $\bq(z)$ in (\ref{e3.2}), and so is also
an adjoint matrix.
With the overall factor of $i$ in $\scon$,
the original contour of
integration for the
constraint field $\lambda(z)$ is along the real axis.

As an aside, note that $\bq(z)$ does not transform
in a simple fashion
under gauge
transformations.  For an infinitesimal gauge transformation $\omega$,
where $A_\mu \rightarrow \partial_\mu \omega - ig [A_\mu, \omega]$,
\beq
\bq(z) \; \rightarrow  \;
\frac{1}{\tvol} \; \int^\beta_0 d\tau
\; \int d^2x_t \; \frac{g c N}{2 \pi T}
\; (- i g) [A_0(x),
\omega(x)] \; ,
\label{e3.5}
\eeq
assuming that $\omega(x)=\omega(\tau,\vxt,z)$ is periodic in $\tau$.
Nevertheless, no further terms besides those in (\ref{e3.3})
are required in the measure of the functional
integral: gauge fixing for $A_\mu$ takes care of that.

The effective action $\seff(\bq)$,
is defined as the integral over all fields, excepting $\bq(z)$:
\beq
Z \speq \int [d\bq(z)] \; e^{- \seff(\bq)} \; .
\label{e3.6}
\eeq
How the effective action is computed in practice
depends upon the problem at hand.  But
by introducing $\bq(z)$ as an extra field into the functional integral,
clearly there is no confusion possible
about double counting degrees of freedom.

The calculation of $\seff(\bq)$ is straightforward at one loop order
[\ref{r7}], and so our discussion is brief.
The field $\bq(z)$ is held fixed, while the gauge field and
the constraint field $\lambda$ are seperated into
classical plus quantum terms:
$A_\mu = \acl_\mu + \aqu_\mu$ and $\lambda = \lcl + \lqu$.
Expanding the action of (\ref{e3.3}) to quadratic order,
\beq
\scl(A) \sppl \sgf(A,\eta) \sppl \scon(\lambda,\bq,A) \; \approx \;
\scal_0 \sppl \scal_1 \sppl \scal_2 \; .
\label{e3.7}
\eeq
The leading term is the sum of the classical action plus the
constraint:
\beq
\scal_0 \speq \scl(\acl) + \scon(\lcl,\bq,\acl) \; .
\label{e3.8}
\eeq

The linear terms determine the equations of motion:
\beq
\scal_1 \speq \scon(\lqu,\bq,\acl) \sppl
 2 \, i \int d^4 x \; tr \left(
\lcl \delta_{\nu 0} \sppl i \dclu_\mu \gcl_{\mu \nu}\right) \aqu_\nu \; .
\label{e3.9}
\eeq
The constraint term, $\scon(\lqu,\bq,\acl)$, is linear in $\lqu$,
and so determines $\acl(x)$.  The obvious choice is to take
$\acl(x)$ to be a function only of $z$,
\beq
\acl(z) \speq \frac{2 \pi T}{g c N} \; \bq(z) \; .
\label{e3.10}
\eeq
The constraint term in $\scal_0$ then
vanishes, $\scon(\lcl,\bq,\acl)=0$.

The second term
in $\scal_1$ modifies the equations of motion for the gauge field:
the constraint term acts as a source for the gauge field, proportional
to $\lcl$.  As $\acl$ is determined, let
\beq
\lcl(z) \speq - i  \dclu_\mu \gcl_{\mu 0}(z) \; ,
\label{e3.11}
\eeq
from which $\scal_1 = 0$.  (As $\acl(x)$ is assumed to depend only upon
$z$, the other components of the equations of motion for the gauge field
are automatically satisfied.)
At the stationary point $\lcl$ is purely imaginary, while
the fluctuations $\lqu$ remain real.  This shift in the contour for
the constraint field $\lambda$ ---
by an imaginary amount, keeping it parallel to the real axis ---
is standard.

The quadratic terms in the action are
$$
\scal_2 \speq
- \; \frac{g c N}{2 \pi T \tvol } \int d^4 x (2i) \; tr\left( \lqu(z)
\;  \aqu_0(x) \right) \;
$$
\beq
\sppl \int d^4 x \int d^4 y \;
tr\left( \aqu_\mu \dic_{\mu \nu} \aqu_\nu
\sppl \bar{\eta} \left( -\dcld^2 \right) \eta \right)\; .
\label{e3.12}
\eeq
The first term in $\scal_2$, involving $\lqu$, is special to the constraint
action.  Since $\lqu$ only enters into $\scal_2$
linearly, it can be integrated
out.  This introduces a constraint for the integration over $\aqu_0(x)$:
\beq
\frac{1}{\tvol} \;
\int^\beta_0 d\tau \; \int d^2x_t \;
\aqu_0(\tau,\vxt,z) \speq 0 \; .
\label{e3.13}
\eeq
This constraint is completely innocuous.
Remember that for the two spatial directions
$\vxt$, each length $L_t$ is taken to infinity.  Integration
is over all modes of $\lqu(z)$ which obey
(\ref{e3.13}), but the only modes which don't are those
constant both in $\tau$ and $\vxt$ --- in momentum space,
modes with $k^0 = \vkt = 0$.
If the length in the $\vxt$ directions are infinite, the corresponding
momenta $\vkt$ take on all continuous values, and those with
$\vkt = 0$ have zero measure, and can be ignored.

Being able to drop the constraint
is an important
point.  Suppose the constraint field is defined not as an integral
over $\tau$ and $\vxt$, but just as an integral over $\tau$: then
$\bq$ is a field in three, instead of one, dimension.
Going through the same proceedure as above, at quadratic order
integration is over all fields constant in time.  But these modes
have {\it nonzero} measure in the functional integral.
At finite temperature, the momentum
$k^0 = 2 \pi n T$ for integral $n$, and the constant modes, with
$n=0$, are of countable extent.

For situations in which the constraint doesn't matter --- that is, where
the effective fields are of zero measure in the space of the original
fields ---
the constraint methods
give the same result as for the usual effective
potential [\ref{r7}].
The constraint field $\lcl$ plays the role of the external
source, while the exchange of $\acl$ for $\bq$ mimics precisely
the process of Legendre transformation.
The effective potential is the energy of the vacuum in the presence
of the external source, and so is properly minimized.

The meaning of the effective action when
the effective fields are of nonzero measure in the space of the original
fields is unclear.  We dwell on this point because of such an
analysis by Oleszczuk and Polonyi [\ref{r10}], who introduce an effective
field in three dimensions, as the integral of $A_0$ with respect to time.
Integrating out modes with $k^0\neq 0$, they find a potential
different from that of (\ref{e2.27}).  The potential in
(\ref{e2.27}) is a constant times $q^2(1-q)^2 = q^2 - 2 q^3 + q^4$.
The potential
for the three dimensional field of [\ref{r10}] is just $q^2 + q^4$:
the term $ -2 q^3$ is missing, as that arises from the
$k^0=0$ mode of the integral.  More generally,
with an effective three dimensional theory,
the manifest \zn symmetry of the original theory
is broken by the separation into modes with zero and nonzero $k^0$.
The loss of the \zn symmetry seems a grievous price to pay.

Returning to (\ref{e3.12}), the remaining terms are standard in a background
field expansion.  In background field gauge, (\ref{e2.15}), the
inverse gauge field propagator is
$$ \dic_{\mu\nu} \speq - \dcld^2 \; \delta^{\mu\nu}
\sppl \dclu_\nu \dclu_\mu \spmn \frac{1}{\xi} \dclu_\mu \dclu_\nu
\sppl i g [\gcl_{\mu\nu},] \;\;\; , $$
\beq
\speq - \dcld^2 \;
\delta^{\mu \nu} \sppl \left( 1 - \frac{1}{\xi} \right)
\dclu_\mu \dclu_\nu \sppl 2 i g [\gcl_{\mu\nu},] \; .
\label{e3.14}
\eeq
After integrating out $\lqu$, $\aqu$, and $\eta$,
\beq
\seff(\bq) \speq \scl(\acl) \sppl \squ_1(\acl) \; ,
\label{e3.15}
\eeq
where $\acl$ is related by $\bq(z)$ by (\ref{e3.10}), and
\beq
\squ_1(\acl) \speq \frac{1}{2} \; tr \; log \left( \dic_{\mu \nu} \right)
\spmn tr \; log\left(- \dcld^2 \right) \; .
\label{e3.16}
\eeq

For the \zn interface let
\beq
\bq(z) \speq q(z) \; t_N \; ,
\label{e3.17}
\eeq
and then repeat the analysis of sec. II.
Taking the field of the \zn instanton as slowly varying,
to leading order in $g^2$ the effective action for
$q(z)$, $\seff(q)$, is given by (\ref{e2.29}).

At next to leading order corrections arise from two sources.
Viewing the action of (\ref{e2.29}) as a type of quantum mechanics,
these terms can be understood as corrections to the
potential and kinetic terms.
As the potential was first generated by the free energy at one
loop order, so corrections to this potential are produced by
the free energy at two loop order.  These two loop effects are
$g^2$ times those at one
loop order.  Like the calculation at
leading order in sec. II,
for the two loop potential the
background field can be taken as constant.
Secondly, at next to leading order it is necessary
to account for the spatial variation of the \zn instanton.
For this only the free energy at one loop order is
required, expanding to leading order in $(dq/dz)^2$;
thus these terms correct the kinetic term in the effective
action.
With our definition of $q$, the classical action
is proportional to $1/g^2$ times $(dq/dz)^2$,
(\ref{e2.14}); the terms from the
free energy are of order one times $(dq/dz)^2$, and so are
smaller by $g^2$.

Ultimately, corrections are small
because the \zn instanton is fat: the ratio of its size to
the thermal wavelength is of order
$1/g$.  Hence an expansion in the derivatives of the instanton field
is automatically an expansion in $g^2$.  This is
what makes the problem tractable.

%!
Both of these effects are due to the effects of
fluctuations in four dimensions as they generate the effective,
one dimensional action $\seff(\bq)$.
The functional integral
over $\bq$ in (\ref{e3.6}), however,
is still treated classically.  When do fluctuations in $\bq(z)$ enter?

Remember that the quantity of physical interest [\ref{r5}]
is the partition
function, $Z$, for a system with the appropriate
boundary conditions to enforce a domain wall in the spacetime tube:
\beq
Z \speq c \; e^{- \, \alpha \tvol} \; .
\label{e3.18}
\eeq
While we have concentrated on the interface
tension, $\alpha$,
of course the prefactor ``$c$'' is also of significance.

Integration over fluctuations in four dimensions
generate the effective
theory in one dimension, $\seff$, and so determine
$\alpha$.  Fluctuations in $\seff$, though, do not contribute to $\alpha$,
only to the prefactor.  This is simply because $\seff$ itself
is proportional to $\tvol$, and so the
integral over the effective, one dimensional fields
cannot generate a constant times $\tvol$ in the exponent, but merely
powers of $\tvol$ in the prefactor [\ref{r5}].
The prefactor for a \zn domain wall is given by the integral
over $\seff$ at one loop order; however, it is necessary to compute
$\seff$ for a general path in group space,
(\ref{ea.1}), instead of the ``classical''
path of (\ref{e3.17}).  This we defer.
%!

Consequently, we confess that
the machinery of the effective action
$\seff(\bq)$ developed in this section
is not essential for what follows.  We discussed
it at such length in order to ensure that
there are no problems of principle,
and to emphasize the generality of the method.

\section{\zn interface at next to leading order: Feynman gauge}

In this section
we compute the leading corrections to the interface tension
in background field
Feynman gauge, $\xi =1$; in the next section, for arbitrary $\xi$.
While the methods are the same for all $\xi$, technically
the calculations are
simpler in Feynman gauge, and so in this section we discuss our
methods in some detail.
In sec. V the results are merely summarized,
in order to emphasize the physical interpretation of the gauge dependence
which arises in the effective action.

As discussed following (\ref{e3.17}),
at next to leading order there are two pieces needed for the interface
tension.  The first is the effective potential in a constant
background $\acl_0$
field to two loop order.  This was calculated for
$SU(3)$ by Belyeav and Eletsky [\ref{r11}] and by
Enqvist and Kajantie [\ref{r12}].  We have independently computed the
potential for the field of (\ref{e2.2}) at arbitrary $N$, but given
previous calculations, are content here to just establish the
$N$--dependence at two loop order.

At one loop order the $N$ dependence of $\squ_1(\acl)$ is obvious.
The background field enters
only through adjoint covariant derivatives;
from (\ref{e2.10}), $t_N$ only has nontrivial commutators with
the generators $t^{\pm}_{N,j}$.  Thus the only $q$--dependence
is from the free energy of the $2(N-1)$ fields for these ladder operators,
and at one loop order $\squ_1(\acl) \sim (N-1)$, (\ref{e2.21}).

The diagrams which enter at two loop order involve either two three--gluon
vertices or one four--gluon vertex.  Both types of diagrams involve
a product of structure constants.
The only diagrams that depend nontrivially upon the background
field are those in which two lines are along
the ladder operators $t^{\pm}_{N,j}$.
Denoting the $q$--dependent terms in the free energy at two loop order
as $\squ_{2}(\acl)$,
after writing each structure constant as a trace
\beq
\squ_2(\acl) \; \sim \;
\sum_{j=1}^{N-1} \sum_{a} \;
\left( tr \left(t^a [t^+_{N,j},t^-_{N,j}]\right) \right)^2 \speq
(N-1) \; \sum_a \left( tr \left(t^a [t^+_{N,(N-1)},
t^-_{N,(N-1)}]\right) \right)^2 \; .
\label{e4.1}
\eeq
The sum is over all generators $t^a$ with nonzero trace.
The last expression follows by noting that from the
form of $t_N$ in
(\ref{e2.3}), each value of ``$j$''
contributes equally to (\ref{e4.1}).  Thus the complete
sum over ``$j$'' is $(N-1)$ times that for any single term,
such as $j=(N-1)$.  The commutator for $t^\pm_{N,(N-1)}$ is:
\beq
[t^+_{N,(N-1)},t^-_{N,(N-1)}] \speq \frac{1}{2} \;
\left(
\begin{array}{ccc}
 & \vdots & \vdots\\
\ldots & -1 & 0 \\
\ldots & 0 & 1 \\
\end{array}
\right) \; .
\label{e4.2}
\eeq
All elements not indicated
vanish.  From (\ref{e4.2}) the only terms
which contribute to (\ref{e4.1}) are if
$t^a$ is one of two diagonal generators,
$t_{N}$ or $t_{N-1}$, (\ref{e2.3}) and (\ref{e4.3}).
It is then easy to show that
the sum over ``$a$'' in (\ref{e4.1})
is proportional to $N$, so that in all
$\squ_2(\acl) \sim N(N-1)$.

Knowing the $N$--dependence, the general result can be read off
from that for $N=3$ [\ref{r11}, \ref{r12}].
For the constant field of (\ref{e2.2}),
the sum of the the free energies at one loop order, $\squ_1(\acl)$
(\ref{e2.27}), and at two loop order in Feynman gauge,
$\squ_{2,\xi=1}(\acl)$, is
$$
\squ_1(\acl) \sppl \squ_{2,\xi=1}(\acl)
$$
\beq
\speq \tvol \; \frac{4 \pi^2 T^4}{3} \; (N-1) \; \int \; dz
\;
\left( q^2 (1-q)^2 \sppl
\left(\frac{g^2 N}{16 \pi^2}\right) \left( 3 \, q^2 (1-q)^2
\spmn 2 \, q(1-q) \right) \right) \; .
\label{e4.4}
\eeq

The second piece required for the interface tension arises from
the free energy at one loop order for a background field which varies
in $z$.  This corrects the kinetic term in the effective action,
as a function of the background field, $q$, times $(dq/dz)^2$:
once one factor of $(dq/dz)^2$ is extracted, the remaining
factors of $q$ can be taken as constant.
At one loop order the quantum action $\squ_1(\acl)$ is given by
(\ref{e3.16}), with the inverse gluon propagator of (\ref{e3.14}),
setting $\xi=1$ in Feynman gauge.  To calculate this
some, although not all, of the tricks of sec. II can be used.
For example, from (\ref{e2.16})-(\ref{e2.18}), for a constant field
the one loop quantum action is independent of $\xi$.  This
is no longer true for a spatially varying field.

The ladder basis of sec. II can be used to simplify the color
algebra.  Defining
$G^+_{0z} = - \, G^+_{z0} =  dq/dz$, else zero,
and $D_+^2 \speq (D^+_0)^2 + \partial_i^2$,
with $D^+_0$ as in (\ref{e2.19}),
for arbitrary fields $q(z)$ the one loop quantum action reduces to
\beq
\squ_{1,\xi =1}(\acl) \speq (N-1) \; \left(
tr \; log \left( -D^2_+ \, \delta_{\mu \nu}
\sppl 4 \pi T \, i \, G^+_{\mu \nu} \right) \spmn
2 \; tr \; log \left( - \, D^2_+\right) \right) \; .
\label{e4.5}
\eeq
One kinetic term arises by expanding
to quadratic order in $G^+_{\mu \nu}$:
\beq
\squ_{1a,\xi=1}(\acl) \speq - \; 16 \pi^2 T^2 \; (N-1)
\; \left(\frac{dq}{dz} \right)^2 \; tr \; \left(
\frac{1}{((k^0_+)^2 + k^2)^2} \right) \; ,
\label{e4.6}
\eeq
with $k_+^0$ as in (\ref{e2.20}).

Having extracted this,
the remaining kinetic term arises from the expansion of
\beq
 \squ_{1b,\xi=1}(\acl)
\speq 2 (N-1) \; \left( 1 \spmn \frac{\epsilon}{2} \right)
\; tr\; log(-D^2_+) \; .
\label{e4.7}
\eeq
The factor of $1-\epsilon/2$ appears
because the theory is regularized
in $4-\epsilon$ dimensions, and in covariant gauges the
number of gluons equals the dimensionality.

Calculating the momentum dependence for such a one loop action
is a standard problem: see, for example, the treatment of
Iliopoulos, Itzykson, and Martin [\ref{r17}].  The computations of
[\ref{r17}], however, are in coordinate space, which
for most problems is rather
awkward.  Instead, it is much simpler to perform the calculations in
momentum space.  We have checked that for the problem at hand,
as well as for the scalar example treated in [\ref{r17}],
the results agree.

To work in momentum space, let $q \rightarrow q + \delta q$
in (\ref{e4.7}), and expand to quadratic order in $\delta q$.
The idea is to isolate the momentum dependence in the fluctuation,
$\delta q$:
\beq
tr \; log \left( -D^2_+ \right)  \; \sim \;
8 \pi^2 T^2 \; \left( \frac{1}{2} \; tr \left(
\frac{1}{-D^2_+} \; (\delta q)^2 \right)
\sppl tr \left( \frac{D^0_+}{-D^2_+} \; \delta q
\; \frac{D^0_+}{-D^2_+} \; \delta q \right) \right) \; .
\label{e4.8}
\eeq
The momentum dependence only arises through the second term on the right
hand side, since the first term is a type of tadpole, independent
of the momentum flowing through $\delta q$.  The field $q$ varies only
in $z$, so its momentum is purely spatial.  If
$(0,\vec{p})$ is the external
momenta, then, and $(k^0,\vec{k})$
the loop momenta, (\ref{e4.8}) becomes
\beq
8  \pi^2 T^2 \; \delta q(\vec{p}) \delta q(-\vec{p})
\; tr \left( \frac{(k^0_+)^2}{((k^0_+)^2 + k^2)
((k^0_+)^2 + (\vec{p} - \vec{k})^2)} \right) \; .
\label{e4.9}
\eeq
In this form it is trivial to expand to order $p^2$.
Trading $L \; (\delta q \, (p^2) \, \delta q)$ for
$\int dz \, (dq/dz)^2$,
the second kinetic term is
$$
\squ_{1b,\xi = 1}(\acl)
$$
\beq
\speq 16 \pi^2 T^2 \; (N-1) \;
\left( \frac{dq}{dz} \right)^2 \;
\left(1 - \frac{\epsilon}{2} \right)
\; tr \left( \frac{- \, k^2}{( (k^0_+)^2 + k^2)^3}
\sppl \frac{4}{3 -\epsilon} \; \frac{k^4}{( (k^0_+)^2 + k^2)^4}
\right) \; .
\label{e4.10}
\eeq
At nonzero temperature dimensional
continuation is carried out
by changing the number of spatial dimensions
to $3-\epsilon$, which produces the factor of $4/(3-\epsilon)$ above.

The integrals required are given in appendix B, (\ref{eb.4}),
(\ref{eb.6}), and (\ref{eb.7}).
Without worrying about their detailed form, one feature is
evident.  Each integral is logarithmically divergent in
four dimensions, so in $4-\epsilon$
dimensions, there are poles in $1/\epsilon$.  These are the
standard terms which produce the renormalization of the coupling
constant at one loop order.  Thus it is instructive
to combine the results
for (\ref{e4.6}) and (\ref{e4.10}) with the classical action of
(\ref{e2.14}) to find
$$
\scl(\acl) \sppl \squ_{1a,\xi=1}(\acl) \sppl \squ_{1b,\xi=1}(\acl)
$$
\beq
\speq \tvol \; \frac{4 \pi^2 T^2}{g^2(T) N} \; (N-1) \;
\int dz \; \left( \frac{dq}{dz} \right)^2
\; \left( 1 \sppl \frac{11}{3} \; \frac{g^2 N}{16 \pi^2}
\left( \psi(q) + \psi(1-q) + \frac{1}{11} \right) \right) \; ;
\label{e4.11}
\eeq
$\psi(q) = d(log \, \Gamma(q))/dq$ is the digamma function.
The prefactor includes the running coupling constant at a
temperature $T$, $g^2(T)$, which is related to the bare coupling
constant $g^2$ as
\beq
\frac{1}{g^2(T)} \speq \frac{1}{g^2} \left(
1 \spmn \frac{11}{3} \; \frac{g^2 N}{16\pi^2} \left(
\frac{2}{\epsilon} \sppl log \left( \frac{\mu^2}{\pi T^2} \right)
\sppl \psi(1/2) \right) \right) \; ,
\label{e4.12}
\eeq
with $\mu$ the renormalization mass scale.
The relationship between the bare and renormalized coupling
constants in (\ref{e4.12}) is arbitrary up to a constant; our choice
is similar but not identical to
the modified minimal subtraction scheme, and
is convenient at nonzero temperature.
At high temperature, (\ref{e4.12}) exhibits
the standard logarithmic
fall off of the running coupling constant, $g^2(T)$,
with the coefficient of $11 N/3$
appropriate for the $\beta$--function of
an $SU(N)$ gauge theory at one loop order.
Notice that
the $q$--dependence in (\ref{e4.11}), through the digamma functions
of $q$ and $1-q$, enters with precisely the same coefficient
as for the $\beta$--function, $11 N/3$.

The effective action which governs the \zn instanton at
next to leading order is the sum of (\ref{e4.4}) and (\ref{e4.12}).
The action is determined by the properties of the solution at
leading order.
As discussed following (\ref{e2.29}), the \zn instanton has zero
energy, and so
$$
\int dz' \; \left( \frac{dq}{dz'} \right)^2
\left( \psi(q) \sppl \psi(1-q) \right)
$$
\beq
\speq 2 \int^{1}_0 dq \; \left( q - \frac{1}{2} \right)
\; log \left( \frac{\Gamma(q)}{\Gamma(1-q)} \right)
\; \sim \;
\spmn .995018\ldots \;\;\; ;
\label{e4.13}
\eeq
the value of the
integral was determined by numerical integration.
Hence at next to leading order, the interface tension
$\alpha$ is
\beq \al \speq \frac{4 (N-1) \pi^2}{3 \sqrt{3N}} \,
\frac{T^3}{g(T)}
\, \left( 1 - (15.2785...)
\frac{g^2(T) N}{16 \pi^2} \right)  \, ,
\label{e4.14}
\eeq
which is the result quoted in (\ref{e1.1}).  We have taken the
liberty of writing the corrections as proportional not just to
the bare coupling constant, $g^2$, but to the running coupling
constant, $g^2(T)$.

\section{\zn interface at next to leading order: general gauges}

The calculation of the interface tension in an arbitrary background
field gauge is similar to that for Feynman gauge.
Nevertheless, the calculation illuminates some
features which are missed by working at fixed $\xi$.

Including the terms at both one and two loop order, the free energy
in the constant background field of (\ref{e2.2}) is
$$
\squ_1(\acl) \sppl \squ_{2,\xi}(\acl)
\speq \tvol \; \frac{4 \pi^2 T^4}{3} \; (N-1) \; \int \; dz \;
\left( q^2 (1-q)^2 \right.
$$
\beq
\left. \sppl \left(\frac{g^2 N}{16 \pi^2}\right) \left(
( 7 \spmn 4 \xi ) \; q^2 (1-q)^2
\spmn ( 3 \spmn \xi )\, q(1-q) \right) \right) \; .
\label{e5.1}
\eeq
The two loop potential for general $\xi$ was computed first by
Enqvist and Kajantie [\ref{r12}]; we agree with their result
when $\xi = 1$, but not for $\xi \neq 1$.

The potential changes in a rather dramatic fashion in going from
one to two loop order.
At one loop order the potential
is just a standard double well, $ q^2(1-q)^2$.  At two loop
order the potential becomes $\xi$--dependent.  This includes
a correction to the coefficient of the double well potential,
as well as a new term, proportional to $q(1-q)$.  This
new term is peculiar, for it controls the behavior of
the potential for small $q$.  If $\xi < 3$, the stable minima
are not at $q=0$ and $q=1$,
but at $q_0 \sim (3 - \xi) g^2$ and
$1-q_0$.  On the other hand, if $\xi > 3$, the stable minima remain
$q=0$ and $q=1$.

Such a nonzero value of the stable minima
would have profound consequences
for a gauge theory at high
temperature [\ref{r9}-\ref{r14}].
For $N \geq 3$, the trace of the
Wilson line in the fundamental representation is a complex number.
Under charge conjugation ($C$) or time reversal ($T$) transformations,
the trace of the
Wilson line goes into its complex conjugate (up to global \zn
transformations).  Thus if the stable vacuum
indeed has $q \neq 0$ (modulo 1), then
the vacuum spontaneously breaks $C$ and $T$ symmetries,
conserving $CT$.  While conceivable,
it is unexpected to find $C$ symmetry breaking arising spontaneously
in a pure gauge theory.  Of course a physical phenomenon cannot
depend upon the choice of the gauge fixing parameter, while
$q_0$ changes with $\xi$.

Leaving these questions aside for the moment,
in a general background gauge the kinetic
terms in the effective action are, to one loop order,
$$
\scl(\acl) \sppl \squ_{1a,\xi}(\acl) \sppl \squ_{1b,\xi}(\acl)
$$
\beq
\speq \tvol \; \frac{4 \pi^2 T^2}{g^2(T) N} \; (N-1) \;
\int dz \; \left( \frac{dq}{dz} \right)^2
\; \left( 1 \sppl \frac{11}{3} \; \frac{g^2 N}{16 \pi^2}
\left( \psi(q) + \psi(1-q) + \frac{7 - 6 \xi}{11} \right) \right) \; .
\label{e5.2}
\eeq
The running coupling constant, $g^2(T)$, remains as in (\ref{e4.12}).

In (\ref{e5.2}) the only $\xi$ dependence is an
overall constant, proportional to $7 - 6 \xi$,
and is independent of the background field $q$.
As discussed following (\ref{eb.7}) in appendix B, for each
of the integrals which contribute to the kinetic term, the
coefficient of the pole in $1/\epsilon$ is always the
same as for
the digamma functions of $q$ and $1-q$, which is how the
$q$--dependence arises.  As is customary
in background field calculations [\ref{r16}],
the poles
in $1/\epsilon$ generate the $\beta$--function at one loop order,
and is independent of $\xi$.  Thus if the coefficient of the
digamma functions is the same as for $1/\epsilon$,
at one loop order it also must be independent of $\xi$.
The only remaining $\xi$--dependence possible is
as a constant, which does appear.

The effective action for arbitrary $\xi$ is the sum of
(\ref{e5.1}) and (\ref{e5.2}).
It is easy to show that while each term
depends individually
upon $\xi$, for any solution with zero energy,
${\cal E} = 0$, the $\xi$--dependence
cancels in the sum.  As the \zn instanton has zero energy,
the value of the interface tension for $\xi \neq 1$ is equal
to that for $\xi =1$, (\ref{e4.14}).

The cancellation of $\xi$--dependence is a necessary check
on the consistency of our method, but by itself is rather
unsatisfactory.  To understand this better
we follow Belyaev [\ref{r14}].
In $SU(2)$ Belyaev showed that the apparent $\xi$--dependence in
the potential for a constant $A_0$ field
can be understood as a renormalization
of the Wilson line.  We now
generalize his results to arbitrary \sun,
and show that they explain the $\xi$--dependence of both the
potential and kinetic terms in the effective action.

The point is that while the vacuum expectation value of the trace
of the Wilson line is a gauge invariant quantity, the fields
which we have been using to parametrize the Wilson line ---
$\acl_0$, and so $q$ --- are not.  For instance, from (\ref{e3.5}),
our effective field transforms in a nonlocal manner under infinitesimal
gauge transformations.
At tree level this doesn't
matter, but it does at one loop order and beyond, as the Wilson
line undergoes both infinite and finite renormalizations.  To
compute these, define
\beq
A_0(x) \speq \acl_0 + \aqu_0(x) \; ,
\label{e5.3}
\eeq
with $\acl_0$ related to $q$ as in (\ref{e2.2}).
In general the Wilson line is a function of the spatial position,
and so we consider the trace of the Wilson line, averaged over
space.  Expanding in powers of $\aqu$,
\beq
\int \frac{d^2x_t}{L^2_t} \;\int \frac{dz}{L}
\; \langle tr \; \Omega(A) \rangle  \speq
tr \; \Omega(\acl) \sppl
\Omega_1 \sppl \Omega_2 \sppl \ldots \; .
\label{e5.4}
\eeq
The first term on the right hand side, $tr \; \Omega(\acl)$, is
the value in the classical
background field, (\ref{e2.4}).
The term linear in $\aqu$, $\Omega_1$, can be written as
\beq
\Omega_1 \speq
ig \; \int \; \frac{dz}{L} \;
\int^\beta_0 d\tau \; \int \frac{d^2x_t}{L^2_t} \;
tr \left(
\; \aqu_0(\tau,\vxt,z) \; \Omega(\acl) \right) \; .
\label{e5.5}
\eeq
Due to the constraint imposed upon the quantum fluctuations
in (\ref{e3.13}), this term vanishes.

The term quadratic in quantum fluctuations is nontrivial.  Consider
first its value in zero background field, $\acl = 0$:
\beq
\Omega_2 \speq
- \; \frac{g^2 (N^2 - 1)}{2} \; \beta \int \; \frac{d^3 k}{(2 \pi)^3}
\; \frac{1}{2 \; k^2} \; .
\label{e5.6}
\eeq
This is a standard renormalization of the Wilson line, a type of
``wave function'' renormalization, proportional to $\beta$,
the length in euclidean time.  In dimensional regularization this
vanishes identically.

However these infinite terms are regularized, they are independent of
the background field.
In addition, there are finite terms which depend upon $\acl$.
Using the commutation relations of (\ref{e2.10}), for an arbitrary
constant ``$y$''
\beq
e^{y t_N} \; t^\pm_{N,j} \; e^{- y t_N} \speq e^{\pm \, y c N} \;
t^\pm_{N,j} \; .
\label{e5.7}
\eeq
Using this relation, and remembering the path ordering required for
the Wilson line, the terms dependent on the background field are
$$
\Omega_2 \speq
- \; \frac{g^2}{N} \; \sum_{j=1}^{N-1} \;
\int^\beta_0 d\tau \int^{\tau'}_0 d\tau' \left(
\aqu_{0,j^+}(\tau) \; \aqu_{0,j^-}(\tau') \;
e^{- 2 \pi i q (\tau - \tau')} \;
tr \left( t^+_{N,j} t^-_{N,j} \; \Omega(\acl) \right) \right.
$$
\beq
\left. \sppl \aqu_{0,j^-}(\tau) \; \aqu_{0,j^+}(\tau') \;
e^{2 \pi i q (\tau - \tau')} \;
tr \left( t^-_{N,j} t^+_{N,j} \; \Omega(\acl) \right) \right) \; .
\label{e5.8}
\eeq
$\aqu_{0,j^\pm}$ is the component of $\aqu_0$ in the
direction of $t^\pm_{N,j}$.  The $\tau$
integrals are evaluated using the background field propagator,
as in (\ref{e2.17}).  After doing the color trace,
in momentum space
\beq
\Omega_2 \speq \frac{i g^2 \beta (N-1)}{2 N} \; e^{2 \pi i q/N}
\; \left(1 \spmn e^{- 2 \pi i q} \right)
\; tr \left( \frac{1}{k^0_+} \; \Delta_{00}(k^0_+,k) \right) \; .
\label{e5.9}
\eeq
$\Delta_{00}(k^0,k)$ is the usual covariant gauge propagator,
\beq
\Delta_{00}(k^0,k) \speq \frac{1}{(k^0)^2 + k^2} \spmn
(1 - \xi) \; \frac{(k^0)^2}{((k^0)^2 + k^2)^2} \; ,
\label{e5.10}
\eeq
with the only dependence on the background field in (\ref{e5.9})
through the shifted momentum $k^0_+$, (\ref{e2.20}).

How then to interpret these corrections at one loop order
to the classical value of the Wilson line?
By the constraint imposed upon $\aqu$, the linear term in (\ref{e5.5})
vanishes, and so they cannot absorbed in $\aqu$.  We then introduce
``renormalized'' fields for $A_0$ and $q$ as
\beq
\aren_0 \speq \frac{2 \pi T}{g c N} \; \qr \; t_N \; \; \; ,
\;\;\; \qr \speq q \sppl \delta q \; ,
\label{e5.11}
\eeq
and require that the spatial average of the vacuum expectation value
of the Wilson line be given by $\aren$:
\beq
\int \frac{d^2x_t}{L^2_t} \; \int \frac{dz}{L} \;
\langle tr \; \Omega(A) \rangle  \speq tr \; \Omega(\aren) \; .
\label{e5.12}
\eeq
Expanding the right hand side to linear order in
$\delta q$, $\delta q$
is proportional to $\Omega_2$.  The integrals required for (\ref{e5.9})
are (\ref{eb.9}) and (\ref{eb.10}) of appendix B, and give
\beq
\qr \speq q \sppl \frac{g^2 N}{16 \pi^2} \; (3 \spmn \xi)
\left( q \spmn \frac{1}{2} \right) \; .
\label{e5.13}
\eeq
This relation is valid for $0 < q < 1$.  For $SU(2)$, it agrees with
the result of Belyaev [\ref{r14}].  Note that the renormalization from
$q$ to $\qr$ is entirely a matter of a finite shift.

The gauge dependent actions, (\ref{e5.1}) and
(\ref{e5.2}), can be trivially rewritten in terms of the renormalized
fields.  The potential term becomes
\beq
\squ_1(\aren) \sppl \squ_{2}(\aren)
\speq \tvol \; \frac{4 \pi^2 T^4}{3} \; (N-1) \; \int \; dz \;
\left( 1 \spmn 5 \;
\frac{g^2 N}{16 \pi^2} \right) \; q^2 (1 \spmn q )^2 \; ,
\label{e5.14}
\eeq
while for the kinetic terms,
$$
\scl(\aren) \sppl \squ_{1a}(\aren) \sppl \squ_{1b}(\aren)
$$
\beq
\speq \tvol \; \frac{4 \pi^2 T^2}{g^2(T) N} \; (N-1) \;
\int dz \; \left( \frac{dq}{dz} \right)^2
\; \left( 1 \sppl \frac{11}{3} \; \frac{g^2 N}{16 \pi^2}
\left( \psi(q) \sppl \psi(1-q) \sppl 1 \right) \right) \; .
\label{e5.15}
\eeq
With the effective action of
(\ref{e5.14}) and (\ref{e5.15}), the
corrections to the interface are of course
unchanged, equal to (\ref{e4.12}).

Once written in
terms of $\qr$, all of the gauge dependence found
previously in the kinetic and potential terms cancels.
Further, the effect of two loop terms in the renormalized potential
is just to change the coefficient of the one loop terms: the
new terms found previously at two loop
order, proportional to $q(1-q)$, have all
been absorbed by $\qr$.

Hence the apparent instability of the perturbative vacuum
at two loop order merely results
from a classical parametrization of the renormalized
Wilson line.  After correcting for loop
effects, the stable vacuum is the trivial one (plus
\zn transforms thereof) and is $C$ symmetric.

It seems unlikely that the cancellations found at two loop
order are mere coincidence.  We conclude with a conjecture:
{\it that the stable vacuum of hot gauge theories --- both
with and without fermions --- is symmetric
under charge conjugation to arbitrary
loop order}.

The research of A.G. and R.D.P. was supported
in part by the U.S. Department of Energy under
contract DE--AC02--76--CH0016.

$\;$\\
\appendixa{\noindent{\bf Appendix A: $\;\;\;$
Proof of minimal action for $N=3$, $\infty$}}
$\;$\\

In this appendix we prove that for $N=3$ and $\infty$,
the path chosen for the \zn instanton is of minimal action.

By a global gauge rotation a constant background field $\acl$ can be
chosen to be a diagonal matrix.  Thus the most general constant field
for the $\acl$ of (\ref{e3.10}) is
\beq
\bq \speq \sum_{i=2}^{N} \; q_i \; t_i \; .
\label{ea.1}
\eeq
The $t_i$ are the $N-1$ diagonal generators of \sun, as in
(\ref{e2.3}) and (\ref{e4.3}).
For clarity, in (\ref{ea.1}) the field $q$ is relabeled as $q_N$.

Promoting each $q_i$ to be a function of $z$, for this ansatz
the classical action becomes a sum over $N-2$ independent
kinetic terms
\beq
\scl(\acl) \speq \tvol \; \frac{4 \pi^2 T^2}{g^2 N}
\; (N-1) \; \int dz   \; \sum_{i=2}^N \;
\left( \frac{dq_i}{dz} \right)^2 \; .
\label{ea.2}
\eeq

For the \zn interface the boundary conditions required are
\beq
q_i(0) \speq q_i(L) \speq 0 \;\;\; , \;\;\; i = 2 \ldots (N-1); \;\;\;\;\;\;
q_N(0) \speq 0 \;\;\; , \;\;\; q_N(L) \speq 1 \; .
\label{ea.3}
\eeq

We wish to show that the path with $q_2(z) = \ldots q_{N-1}(z) = 0$,
and $q_N(z) = q(z)$ as before, is the path of minimal action.
To demonstrate this we need the potential generated by fluctuations
at one loop order.  We do so in two special cases where the analysis
is elementary.

$\;$\\
\noindent{\it N=3:}
There are two independent fields, $q_2$ and $q_3$.  The potential
term is
$$
\squ_1(\acl) \speq \tvol \; \frac{4 \pi^2 T^4}{3} \;
\int dz \;  \cv_{tot}(q_2,q_3) \; ,
$$
\beq
\cv_{tot}(q_2,q_3) \speq  \cv(q_2) \sppl
\cv(q_2/2 + q_3) \sppl \cv(-q_2/2 + q_3)  \; ,
\label{ea.4}
\eeq
where
\beq
\cv(q) \speq [q]^2 (1 - [q])^2 \;\;\; , \;\;\;
[q] \speq |q|_{mod \, 1} \; .
\label{ea.5}
\eeq
Because the $\cv(q)$ is a function of $[q]$, the absolute
value modulo one, it is not quite the simple polynomial form it first
appears to be.  (This restriction could be ignored before, since the
\zn instanton only involves $q: 0 \rightarrow 1$.)

The effective action is the sum of (\ref{ea.3}) and (\ref{ea.4}).
It is not difficult to see why for the boundary conditions of
(\ref{ea.3}), the path with $q_2(z)=0$ is of minimal action.
Considering the problem as classical mechanics in two
dimensions, from energy conservation
the action for any solution to the equations
of motion is proprotional to
\beq
\int ds \; \sqrt{\cv_{tot}(q_2,q_3)} \; ,
\label{ea.6}
\eeq
where $ds$ is the arc length in the space of $q_2$ and $q_3$.
It can be shown that for any {\it fixed}
value of $q_3$, the potential is minimized for $q_2 = 0$:
$\cv_{tot}(q_2,q_3) \geq \cv_{tot}(0,q_3)$.
Given the boundary conditions, the path chosen is clearly of minimal
length.  As both the arc length and the potential are bounded by
the path with $q_2 = 0$, by (\ref{ea.6}), so is the action.

$\;$\\
\noindent{$\it N=\infty$:}
The large $N$ limit is taken by holding
$\widetilde{g}^2 \equiv g^2 N$ fixed as $N \rightarrow
\infty$; we work in weak coupling, for small $\widetilde{g}$.
Then the interface tension is of order $N$,
$\alpha \sim N/\widetilde{g}$,
while the \zn instanton remains fat, with a width of order
$1/(\widetilde{g} T)$.

Besides $q_N$, there are of order $N$ $q_i$'s which can contribute.
Assume first that there are a finite fraction of the $q_i$'s
for which $q_i(z) \neq 0$.  From (\ref{ea.3}),
since each field has a kinetic term proportional
to $N$, if order $N$ fields contribute, the sum of the kinetic terms
for all fields is of order $N^2$.  Similarly, the
potential term is also of order $N^2$.
Because both the kinetic
and potential terms are positive definite, any solution has positive
action, equal to
a pure number times $N^2$.  This is $N$ times
the action of the path chosen and so is not minimal.

Hence we can assume that at large $N$, only a {\it finite} number of
the $q_i$'s contribute.
For simplicity, assume that they are just
$q_{N}$ and $q_{N-1}$.   The generators for $t_{N}$,
(\ref{e2.3}), and $t_{N-1}$, (\ref{e4.3}), simplify greatly at large $N$,
reducing to just one diagonal element.  (Essentially, the terms which
enforce tracelessness can be ignored, as a correction in $1/N$.)
In this case it is easy to work out the full potential term,
which is proportional to
$$
\cv_{tot}(q_{N-1},q_N) \speq
(N-2) \cv(q_{N}) \sppl (N-2) \cv(q_{N-1}) \sppl \cv(q_{N-1} - q_{N})
$$
\beq
\; \sim \; N \left( \cv(q_N) \sppl \cv(q_{N-1}) \right) \; .
\label{ea.7}
\eeq
The last term is the leading term
at large $N$; like the kinetic energy, it is of order $N$.
But notice that at large $N$, in the potential
the coupling between $q_{N}$ and $q_{N-1}$
has dropped out.
Thus even though the potential for $q_{N-1}$ is
nontrivial,
because it is positive definite, the path of minimal
action which satisfies $q_{N-1}(0) = q_{N-1}(L) = 0$ is just
$q_{N-1}(z) = 0$.

This argument generalizes immediately to any finite number of $q_i$'s:
the term of order $N$ in the potential is a sum over decoupled
potentials, and the path of minimal action excites only one field,
$q_N$.

$\;$\\
%\newpage
\appendixb{\noindent{\bf Appendix B: $\;\;\;$
Integrals in a constant $A_0$ field}}
$\;$\\

In this appendix we catalog the integrals required for the computation
of quantum actions in the constant background field of (\ref{e2.2}).

The trace is defined as
\beq
tr \speq V \;T \sum_{n = - \infty}^{+\infty}
\; \int \frac{d^{3-\epsilon}k}{(2 \pi)^{3 - \epsilon}}
\; .
\label{eb.1}
\eeq
Remember that the system is of length $L_t$ in the $\vec{x}_t$ directions,
and $L$ in the $z$ direction.
At a temperature $T$, $\beta = 1/T$,
and the momentum $k^0 = 2 \pi n T$ for integral $n$.
The volume of spacetime $V = \beta L^2_t L$.
Dimensional continuation in
$3-\epsilon$ dimensions is used.

As discussed following (\ref{e2.21}),
after using the ladder basis to reduce the
color algebra, the propagators in the background field of (\ref{e2.2})
are given by their values in zero field, except for
the replacement of $k^0 \rightarrow
k^0_+ = 2 \pi T (n + q)$.  Following the example of
(\ref{e2.21})--(\ref{e2.27}), the integrals are most easily
computed by zeta function techniques [\ref{r15}].
First, the spatial integrals
over $d^{3-\epsilon}k$ are done using the standard formulas of dimensional
regularization.  This leaves a sum over $n$, which is evaluated
in terms of zeta functions.
For instance, the simplest integral which arises in the
expansion of the kinetic term at one loop order is
\beq
tr \; \frac{1}{((k^0_+)^2 + k^2)^2} \speq
\frac{V}{16\pi^2} \left( 1 \sppl \frac{\epsilon}{2} \right) \;
\left(\psi(1/2) \spmn log(\pi T^2) \right)
\; \sum_{n=-\infty}^{+\infty} \frac{1}{|n+q|^{1+\epsilon}} \; .
\label{eb.2}
\eeq
Using the definition of the zeta function, (\ref{e2.24}),
and then expanding to order $\epsilon$,
\beq
\sum_{n=-\infty}^{+\infty} \frac{1}{|n+q|^{1+\epsilon}}
\speq \zeta(1+\epsilon,q) \sppl \zeta(1+\epsilon,1-q)
\; \sim \; \frac{2}{\epsilon} \spmn
\left( \psi(q) \sppl \psi(1-q) \right)	\; ,
\label{eb.3}
\eeq
with $\psi(q) = d(log\, \Gamma(q))/dq$ the digamma function.  Hence in all
\beq
tr \;
\frac{1}{((k^0_+)^2 + k^2)^2} \speq
\frac{V}{16\pi^2} \;
\left( \frac{2}{\epsilon'} \spmn
\left( \psi(q) \sppl \psi(1-q) \right)
\right)	\; ,
\label{eb.4}
\eeq
where
\beq
\frac{2}{\epsilon'} \speq \frac{2}{\epsilon}
\sppl \psi(1/2) \spmn log(\pi T^2) \; .
\label{eb.5}
\eeq
Similarly,
\beq
tr \;
\frac{(k^0_+)^2}{((k^0_+)^2 + k^2)^3} \speq
\frac{V}{64\pi^2} \;
\left( \frac{2}{\epsilon'} \spmn
\left( \psi(q) \sppl \psi(1-q) \right)
\sppl 2 \right)	\; ,
\label{eb.6}
\eeq
\beq
tr \;
\frac{(k^0_+)^4}{ ( (k^0_+)^2 + k^2 )^4} \speq
\frac{V}{128\pi^2} \;
\left( \frac{2}{\epsilon'} \spmn
\left( \psi(q) \sppl \psi(1-q) \right)
\sppl \frac{8}{3} \right) \; ,
\label{eb.7}
\eeq

These integrals are all those required to compute the one loop corrections
to the kinetic term in sec.'s IV and V.  The factors of the renormalization
group scale $\mu$ which enter into the running coupling constant,
(\ref{e4.12}), arise by taking $g^2 \rightarrow g^2 \mu^{\epsilon}$
in the classical action, and expanding in $\epsilon$.

Notice that for the three integrals of
(\ref{eb.4}), (\ref{eb.6}), and (\ref{eb.7}),
the dependence on the background field $q$,
through the digamma functions, always has the same
coefficient as the pole in $1/\epsilon$.
This is because in all of these integrals, $q$ enters
through the
zeta functions $\zeta(1+\epsilon,q) +
\zeta(1+\epsilon,1-q)$.  Expanding
about small $\epsilon$, (\ref{eb.3}),
generates a common factor of $1/\epsilon$
and digamma functions in each case, so that
up to their overall normalization,
these integrals only differ by a constant.

Other integrals required for the potential at two loop order, and the
renormalization of the Wilson line at one loop order, include
(\ref{e2.21}), (\ref{e2.25}), and the following.  These are all finite
integrals, so it is safe to set $\epsilon = 0$.
\beq
tr \;
\frac{1}{(k^0_+)^2 + k^2} \speq
\frac{V \, T^2}{2} \; \left( q^2 \spmn q \sppl \frac{1}{6} \right) \; ,
\label{eb.8}
\eeq
\beq
tr \;
\frac{1}{k^0_+ \; ((k^0_+)^2 + k^2)} \speq
\frac{V \, T}{2 \pi} \; \left( q \spmn \frac{1}{2} \right) \; ,
\label{eb.9}
\eeq
\beq
tr \; \frac{k^0_+}{((k^0_+)^2 + k^2)^2} \speq
- \; \frac{V \, T}{4 \pi} \; \left( q \spmn \frac{1}{2} \right) \; .
\label{eb.10}
\eeq
The last two integrals, (\ref{eb.9}) and (\ref{eb.10}), are
singular for $q=0$ or $1$, and as written are defined over
$0 < q < 1$.  Properly defined, they are discontinuous and vanish for
$q=0$ or $1$.

\newpage
\noindent{ \bf References}
\newcounter{nom}
\begin{list}{[\arabic{nom}]}{\usecounter{nom}}
\item
G. 't Hooft, \nup{B138}{1}{78}.
\label{r1}
\item
B. Svetitsky and L. G. Yaffe, \nup{B210}{423}{82}.
\label{r2}
\item
K. Kajantie, L. K\"{a}rkk\"{a}inen, \phl{B214}{595}{88};
K. Kajantie, L.  K\"{a}rkk\"{a}inen, K. Rummukainen, \nup{B333}{100}{90};
K. Kajantie, L.  K\"{a}rkk\"{a}inen, K. Rummukainen, \nup{B357}{693}{91};
S. Huang, J. Potvin, C. Rebbi and S. Sanielevici, \prd{42}{2864}{90},
(E) \prd{43}{2056}{91};
J. Potvin and C. Rebbi, Boston University preprint 91-0069 (Jan. 1991);
R. Brower, S. Huang, J. Potvin, C. Rebbi, and J. Ross, Boston University
preprint 91-22 (Nov. 1991);
R. Brower, S. Huang, J. Potvin, and C. Rebbi, Boston University preprint
92-3 (Jan. 1992).
\label{r3}
\item
T. Bhattacharya, A. Gocksch, C. P. Korthals Altes and R. D. Pisarski,
\prl{66}{998}{91}.
\label{r4}
\item
E. Br\'{e}zin and J. Zinn-Justin, \nup{B257 [FS14]}{867}{85}.
\label{r5}
\item
G. M\"{u}nster, \nup{B324}{630}{89}; \nup{B340}{559}{90}.
\label{r6}
\item
R. Fukuda and E. Kyriakopoulos, \nup{B85}{354}{354};
L. O'Raifeartaigh, A. Wipf, and H. Yoneyama, \nup{B271}{653}{86};
C. Wetterich, \nup{B352}{529}{91}.
\label{r7}
\item
D. Gross, R.D. Pisarski, L.G. Yaffe, \rmp{53}{43}{81};
N. Weiss, \prd{24}{75}{81}; \prd{25}{2667}{82};
\label{r8}
\item
R. Anishetty, \jph{G10}{439}{84};
K. J. Dahlem, \zpc{C29}{553}{85};
S. Nadkarni, \prl{60}{491}{88};
\label{r9}
\item
J. Polonyi, \nup{A461}{279}{87};
J. Polonyi and S. Vazquez; \phl{B240}{183}{90};
M. Oleszczuk and J. Polonyi, MIT preprint MIT-CTP-1984 (June, 1991).
\label{r10}
\item
V. M. Belyaev and V. L. Eletsky, \zpc{45}{355}{90}.
\label{r11}
\item
K. Enqvist and K. Kajantie, \zpc{47}{291}{90}.
\label{r12}
\item
V. M. Belyaev, \phl{B241}{91}{90}.
\label{r13}
\item
V. M. Belyaev, \phl{B254}{153}{91}.
\label{r14}
\item
A. Actor, \nup{B265 [FS15]}{689}{86}; \fph{35}{793}{87};
H. A. Weldon, \nup{B270}{79}{86};
R. V. Konoplich, \tmp{78}{315}{89};
F. T. Brandt, J. Frenkel, and J. C. Taylor, \prd{44}{1801}{91}.
\label{r15}
\item
G. 't Hooft and M. Veltman, \nup{B44}{189}{72};
L. F. Abott, \nup{B185}{189}{81}.
\label{r16}
\item
J. Iliopoulos, C. Itzykson, and A. Martin,
\rmp{47}{165}{75}.
\label{r17}
\end{list}
\end{document}